\documentclass[12pt, onecolumn]{IEEEtran}

\makeatletter
\def\ifundefined{\@ifundefined}
\makeatother

\usepackage{setspace}
\usepackage{srcltx}  
\usepackage{amsmath}
\usepackage{amssymb}
\usepackage{graphics}
\usepackage{cite}
\usepackage{psfrag}
\usepackage{epsfig}
\usepackage{subfigure}
\usepackage[margin=2.5cm]{geometry}

\newtheorem{thm}{Theorem}
\newtheorem{cor}{Corollary}
\newtheorem{fact}{Fact}

\doublespace

\begin{document}
\renewcommand{\textfraction}{0}

\title{On Finite-Length Performance of Polar Codes: Stopping Sets, Error Floor, and Concatenated Design}


\author{A. Eslami and H. Pishro-Nik
\thanks{The material in this paper was presented in part at International Symposium of Information Theory (ISIT), 2011, and 48th Annual Allerton Conference on Communication, Control, and Computing, 2010. This work was supported by National Science
Foundation under grants CCF-0830614 and ECCS-0636569.}
\thanks{The authors are with the
Electrical and Computer Engineering Department, University of Massachusetts, Amherst, MA, USA ~(email:eslami, pishro@ecs.umass.edu).}}

\maketitle
\thispagestyle{empty}
\vspace{-.4 in}
\begin{abstract}
\vspace{-.1 in}
This paper investigates properties of polar codes that can be potentially useful in real-world applications.
We start with analyzing the performance of finite-length polar codes over the binary erasure channel (BEC), while assuming belief propagation as the decoding method. We provide a stopping set analysis for the factor graph of polar codes, where we find the size of the minimum stopping set. We also find the girth of the graph for polar codes.
Our analysis along with bit error rate (BER) simulations demonstrate that finite-length polar codes show superior error floor performance compared to the conventional capacity-approaching coding techniques.
In order to take advantage from this property while avoiding the shortcomings of polar codes, we consider the idea of combining polar codes with other coding schemes.
We propose a polar code-based concatenated scheme to be used in \emph{Optical Transport Networks} (OTNs) as a potential real-world application. Comparing against conventional concatenation techniques for OTNs, we show that the proposed scheme outperforms the existing methods by closing the gap to the capacity while avoiding error floor, and maintaining a low complexity at the same time.

\end{abstract}

\normalsize
\vspace{-.1 in}

\begin{keywords}
\vspace{-.1 in}
Polar Codes, Concatenated Codes, Belief Propagation, Stopping Sets, Error Floor.
\end{keywords}

\vspace{-.1 in}

\section{Introduction}
Since their introduction, polar codes have attracted a lot of attention among researchers due to their capability to solve some problems (sometimes open problems) that could not be handled using other schemes.
However, theoretical approaches have been mostly taken toward polar codes in the literature. Our goal is to study polar codes from a practical point of view to find out about properties that can be useful in real-world applications. Hence, we are mainly concerned with the performance of polar codes in the finite regime (i.e. with finite lengths)  as opposed to the asymptotic case. Some of the previous work related to finite-length polar codes include \cite{moriitw10, Tallist11, Seidl10, Alamdar11, goelaitw10, Pedarsani11, Talhowto11}. Particularly, \cite{Tallist11} proposes a successive cancellation list decoder that bridges the gap between successive cancellation and maximum-likelihood decoding of polar codes. Inspired by \cite{Tallist11}, \cite{Bonik12,BinLi12,Niu12} propose using CRC along with list decoding to improve the performance of polar codes.
\cite{Seidl10} presents a method to improve the finite-length  performance of successive cancellation decoding by means of simple and short inner block codes. A linear program (LP) decoding for polar codes is considered in \cite{goelaitw10}. In \cite{Talhowto11}, a method for efficient construction of polar codes is presented and analyzed. In addition, scaling laws are provided in \cite{hassani10, hassani102, koradaisit10, Goli12, Hassaniratedep11} for the behavior of polar codes that, in some cases, have finite-length implications.

Since an analysis in the finite regime can be very difficult in general,  we start with studying the performance of polar codes over the binary erasure channel (BEC). While being fairly manageable, such an analysis leads to a better understanding of the behavior of polar codes.
We provide an analysis of the stopping sets in the \emph{factor graph realization} of polar codes. Such a realization for polar codes was first employed by \cite{arikan08} and \cite{hussami09} to run Belief Propagation (BP) as the decoding algorithm.
Stopping sets are important as they contribute to the decoding failure and error floor, when BP is used for decoding \cite{Urbankefinite02}.
Particularly, in the case of BEC, stopping sets are the sole reason of the decoding failure.
We find the structure of the minimum stopping set and its size, called \emph{stopping distance}.
We will show that the stopping distance grows polynomially for polar codes. This is a clear advantage over capacity-approaching LDPC codes. We also find the girth of the factor graph of polar codes, showing that polar codes hold a relatively large girth. The effect of such a large girth and stopping distance on the error floor behavior of polar codes is depicted in our simulation results for the binary erasure and AWGN (Additive White Gaussian Noise) channels.

It is well-known that finite-length polar codes show poor error probability performance when compared to some of the existing coding schemes such as LDPC and Turbo codes. Nevertheless, showing a set of good characteristics such as being capacity-achieving, low encoding and decoding complexity, and good error floor performance suggests that a combination of polar coding with another coding scheme could eliminate shortcomings of both, hence providing a powerful coding paradigm.
In this paper, we consider the design of polar code-based concatenated coding schemes that can contribute to closing the gap to the capacity. Concatenated coding has been studied extensively for different combinations
of coding schemes. Furthermore, there
have been many applications, such as deep space communications, magnetic recording channels, and optical transport systems that use a concatenated coding scheme \cite{kurtas06, wu10, ningde08, Mizuochi09}.
A coding scheme employed in these applications needs to show strong error correction capability.
Here, we investigate the potentials of using polar codes in a concatenated scheme to achieve very low error rates while avoiding error floor.
While the idea of concatenated polar codes was first introduced in \cite{bakshi10}, the problem of designing practical concatenated schemes using polar codes is yet to be studied. In \cite{bakshi10}, the authors study the classical idea of
code concatenation using short polar codes as inner codes and a high-rate Reed-Solomon (RS) code
as the outer code. It is shown that such a concatenation scheme with a careful choice of parameters
boosts the rate of decay of error probability to almost exponential in the block-length with
essentially no loss in computational complexity. While \cite{bakshi10} mainly considers the asymptotic case, we are interested in improving the performance in practical finite lengths.

In this paper, we study the combination of polar codes and LDPC codes, suggesting a polar code as the outer code and a LDPC code as the inner code.
LDPC codes can be decoded in linear time using BP, while they can get very close to the capacity. However, LDPC codes with good waterfall characteristics are known to mostly suffer from the error floor problem. Here, polar codes come to play their role making the combination to show a good error floor performance.
In order to investigate the performance of this scheme in a real-world application, we compare our proposed scheme against some of the conventional schemes used for OTNs. These schemes include a capacity-approaching LDPC code, the ITU-T Recommendation G.709 for OTNs, and some of the ``super codes" of ITU-T G.975.1 for DWDM (Dense Wavelength Division Multiplexing) submarine cable systems. we will show that polar-LDPC combination actually outperforms these schemes as it closes the gap to capacity without showing error floor. Our results suggest that polar codes have a great potential to be used in combination with other codes in real-world communication systems.

The rest of the paper is organized as follows. We first explain the notations and provide a short background on the belief propagation. Section \ref{sec:stopset} gives an analysis on the minimum stopping set of polar codes. We provide a girth analysis of polar codes in Section \ref{sec:errorfloor} where we also present simulation results for error floor performance. We propose concatenated polar codes to be used in a real-world application in Section \ref{sec:concat}. Finally, Section \ref{sec:conclusion} concludes the paper.

\section{Preliminaries}\label{sec:BP}
In this section, we explain the notations and some preliminary concepts we will use in our analysis.
Let $F = \left[ \begin{smallmatrix} 1&0\\ 1&1 \end{smallmatrix} \right]$ be the kernel used for construction of polar codes.
Apply the transform $F^{\otimes n}$ (where $\otimes n$ denotes the $n$th Kronecker power) to a block
of $N = 2^n$ bits and transmit the output through independent
copies of a symmetric \emph{binary discrete memoryless channel (B-DMC)}, call it $W$. As $n$ grows large,
the channels seen by individual bits (suitably defined in \cite{arikan09})
start polarizing to either a noiseless channel or a
pure-noise channel, where the fraction of channels becoming
noiseless is close to the capacity $I(W)$.
Polar codes use the noiseless channels for transmitting information while fixing
the symbols transmitted through the noisy ones to a value
known both to the sender as well as the receiver.
Accordingly, part of the block that carries information includes ``information bits" while the rest of the block includes ``frozen bits".
Since we only deal with symmetric channels in this paper, we assume without loss of generality that the fixed positions are set to 0.
The code is defined through its generator matrix as follows. Compute the
Kronecker product $F^{\otimes n}$. This gives a $2^n \times  2^n$ matrix. The generator matrix of polar codes is a sub-matrix of $F^{\otimes n}$ in which only a subset of rows of $F^{\otimes n}$ are present. These rows are in fact the rows of $F^{\otimes n}$ corresponding to  information bits. In the following, let $\bar{x}=(x_1, ..., x_N)$ and $\bar{y}=(y_1, ..., y_N)$ denote, respectively, the vectors of code-bits and channel output bits.

A Successive Cancelation (SC) decoding scheme is employed in \cite{arikan09} to prove the capacity-achieving property of polar codes. However, \cite{arikan08} and \cite{hussami09} later proposed using belief propagation decoding to obtain better BER performance while keeping the decoding complexity at $O(N \log N)$. Belief propagation can be run on the factor graph representation of the code \cite{arikan08}. Such a representation is easily obtained by adding check nodes to the encoding graph of polar codes, as it is shown in Fig. \ref{fig:stopset1} for a code of length 8. We refer to this graph as the code's \emph{factor graph}. Note that the factor graph is formed of columns of variable nodes and check nodes.  There are, respectively, $n+1$ and $n$ columns of variable and check nodes in the graph.
We denote the variable nodes in $j$th column  by $v(1,j), v(2,j), ..., v(N,j)$ for $j=1, ..., n+1$. This is also shown in Fig. \ref{fig:stopset1}. Similarly, check nodes are labeled as $c(1,j), c(2,j), ..., c(N,j)$ for $j=1, ..., n$. The rightmost column in the graph includes code-bits, while the leftmost column includes frozen and information bits. As it will become clear, our analysis does not depend on any specific choice of the frozen and information bits. Therefore, we treat all the nodes in the left-most column as variable nodes. Among $v(i,1), i=1, ..., N$, some are associated to the information bits. We denote the index set of information bits by $\mathcal{A}$ where $\mathcal{A}\subseteq \{1,2,...,N\}$. Also, the row in $F^{\otimes n}$ associated with an information bit $i\in \mathcal{A}$ will be denoted by $\textbf{r}_i= [r_{i,1}\  r_{i,2}\  ...\  r_{i,N}]$. Note that this is the $i$th row of $F^{\otimes n}$. We denote by $wt(\textbf{r}_i)$ the Hamming weight of $\textbf{r}_i$.


BP runs on the factor graph in a column-by-column fashion. That is, BP runs on each column of the adjacent variable and check nodes. The parameters are then passed to the next column. Each column, as it can be seen in Fig. \ref{fig:stopset1}, is formed of some Z-shaped subgraphs. In our proofs, we sometimes simply call a Z-shaped part a ``Z". The schedule with which BP runs is very important for channels other than BEC. Here, we use the same scheduling used in \cite{hussami09}, i.e. we update the LLRs for Z parts from bottom to top for each column, starting from the rightmost one. After arriving at the leftmost column, we reverse the course and update the Zs from top to bottom for each column, moving toward the rightmost one. This makes one round of iteration, and will repeat at each round. While we tried other schedules as well, this one led to a better overall performance.

We denote the factor graph of a code of length $N=2^n$ by $T_n$. A key observation is the symmetric structure of this graph due to the recursive way of finding the generator matrix: $T_{n+1}$ includes two factor graphs $T_n$ as its upper and lower halves, connected together via $v(1,1), v(2,1), ..., v(N,1)$ and $c(1,1), c(2,1), ..., c(N,1)$. We denote these two subgraphs by $T_{n+1}^{U}$ and $T_{n+1}^{L}$, as it is shown in Fig. \ref{fig:stopset1}. This observation will be later used in our analysis.

In this paper, we are particularly interested in the analysis of \emph{stopping sets} in the factor graph of polar codes. A stopping set is a non-empty set of variable nodes such that every neighboring check node of the set is connected to at least two variable nodes in the set.
Fig. \ref{fig:stopset1} shows an example of the stopping set in the polar codes' graph, where we have also included the corresponding set of check nodes.
A stopping set with minimum number of variable nodes is called a \emph{minimum stopping set}.

\subsection{Stopping Trees}
An important category of stopping sets in the factor graph of polar codes are \emph{stopping trees}.
A stopping tree is a stopping set that contains one and only one information bit. It can be easily seen that this sub-graph is indeed a tree, therefore justifying its name. We say that the stopping tree is rooted at its (single) information bit (on the left side of the graph), with leaves at code-bits (on the right side of the graph).
An example of such a stopping set is shown in Fig. \ref{fig:stopset2} with black variable nodes. We also included the corresponding set of check nodes in order to visualize the structure of the tree. A stopping tree like the one shown in Fig. \ref{fig:stopset2} can be immediately realized for any information bit.
As we will later see (in Fact \ref{fact:uniqstoptree} below), this would in fact be the unique stopping tree for each information bit.
We denote the stopping tree rooted at $v(i,1)$ by $ST(i)$.
Among all the stopping trees, the one with minimum number of variable nodes is called a \emph{minimum stopping tree}.
We refer to the set of leaf nodes of a stopping tree as the \emph{leaf set} of the tree. The size of the leaf set for $ST(i)$ is denoted by $f(i)$. We refer to a stopping tree with minimum leaf set as a \emph{Minimum-Leaf Stopping Tree (MLST)}.
Note that a minimum stopping tree does not necessarily have the minimum $f(i)$ among all the stopping trees.

\subsection{Graph Stopping Sets vs. Variable-Node Stopping Sets}
By looking at the factor graph of polar codes, one can observe that the middle variable nodes, i.e. $v(i,j)$ for $j=2, ...,n$ and $i=1, ...,N$, are always treated as erasures by the BP decoder. This is also true about information bits. Frozen bits, on the other hand, are known to the decoder.
As a result, the only real ``variable" nodes are the code-bits, i.e. $v(1,n+1), ...,v(N,n+1)$. These are in effect the variable nodes that if erased may cause a decoding failure.  Here, we refer to a stopping set on the graph as a \emph{Graph Stopping Set (GSS)}, while we refer to the set of code-bits on such a GSS as a \emph{Variable-Node Stopping Set (VSS)}. In Fig. \ref{fig:stopset1}, the set $\{x_3, x_4, x_5, x_6\}$ is the VSS for the depicted GSS.
As we will see later, every GSS must include some information bits and some code-bits. Thus, VSS is nonempty for each GSS.
Accordingly, we define a \emph{minimum VSS} (\emph{MVSS}) as a VSS with minimum number of code-bits among all the VSSs. That is, a minimum VSS is the set of code-bits on a GSS with minimum number of code-bits among all GSSs. Note that a minimum VSS is not necessarily on a minimum GSS. We refer to the size of a minimum VSS as \emph{stopping distance} of the code.

Now, for any given index set $J\subseteq \mathcal{A}$, there always exists an information bit $j\in J$ whose corresponding stopping tree has the smallest leaf set among all the elements in $J$. We call such an information bit a \emph{minimum information bit} for $J$, denoted by $MIB(J)$. Note that there may exist more than one MIB in $J$.
In general, any given index set $J\subseteq \mathcal{A}$ can be associated to several GSSs in the factor graph. We denote by $GSS(J)$ the set of all the GSSs that include $J$ and only $J$ as information bits.
Each member of $GSS(J)$ includes a set of code-bits. The set of code-bits in each of these GSSs is a VSS for $J$.
We refer to the set of these VSSs as \emph{variable-node stopping sets} (VSSs) of $J$, denoted by $VSS(J)$.  Among the sets in $VSS(J)$, we refer to the one with minimum cardinality as a \emph{minimum VSS for $J$}, denoted by $MVSS(J)$.
Let us also mention that all the proofs for the facts, lemmas, and theorems have been moved to the Appendix at the end of the paper.

\section{Stopping Set Analysis of Polar Codes} \label{sec:stopset}
In this section, we provide a stopping set analysis for polar codes.
For the BEC, it is proved \cite{Urbankefinite02} that the set of erasures which remain when the decoder stops is equal to the unique maximal stopping set within the erased bits.
In general, an analysis of the structure and size of the stopping sets can reveal important information about the error correction capability of the code. A minimum stopping set is generally more likely to be erased than larger stopping sets. Thus, minimum stopping sets play an important role in the decoding failure. In code design, codes with large minimum stopping sets are generally desired.
We consider the problem of finding the minimum stopping set for a given polar code of length $N$. The results of this analysis may also help finding the optimal rule of choosing information bits to achieve the best error correction performance under belief propagation decoding.

\subsection{Minimum VSS in The Graph}\label{sec:MVSS}
It is important though to realize that what prevents the BP decoder from recovering a subset $J$ of information bits is the erasure of the code-bits in one of the sets in $VSS(J)$. Therefore, what will eventually show up in any error probability analysis is the set of VSSs and their sizes.
Particularly, $MVSS(J)$ represents the smallest set of code-bits whose erasure causes a decoding failure of $J$. We will find the size of $MVSS(J)$ for any given $J$. Furthermore, we will find the size of minimum VSS for a given polar code.

We start our analysis by stating some of the facts about the structure of stopping sets in the factor graph of polar codes.
The factor graph of polar codes has a simple recursive structure which points to some useful observations.
Here we mention some of these observations.
\begin{fact}\label{fact:stopends}
Any GSS in the factor graph of a polar code includes variable nodes from all columns of the graph. In particular, any GSS includes at least one information bit and one code-bit. $\blacksquare$
\end{fact}

This implies that any given GSS includes a nonempty VSS.

\begin{fact}\label{fact:uniqstoptree}
Each information bit has a unique stopping tree. $\blacksquare$
\end{fact}

\begin{fact}\label{fact:stopinduce}
Any GSS in $T_{n+1}$ is formed of a GSS in $T_{n+1}^{U}$ and/or a GSS in $T_{n+1}^{L}$, and a number of variable nodes $v(i,1), \ i=1, ..., N$. $\blacksquare$
\end{fact}

This implies that any GSS in $T_{n+1}$ induces a GSS in $T_{n+1}^{U}$ and/or $T_{n+1}^{L}$. This can be also seen in Fig. \ref{fig:stopset1}.
The stopping set shown in the figure induces a stopping set in each of $T_{n+1}^{U}$ and $T_{n+1}^{L}$. Now, consider size of the leaf set for different stopping trees.
Note that we have $f(1)=1$, $f(2)=2$, $f(3)=2$, $f(4)=4$, so on. In general, we can state the following facts about $f(\cdot)$.
\begin{fact}\label{fact:f}
For a polar code of length $N=2^n$, the function $f(\cdot)$ can be formulated as follows:
\begin{align}
\nonumber &f(2^l)=2^l  \quad  \text{for} \quad l=0, 1, ...,n, \\
 &f(2^l+m)=2f(m) \quad \text{for} \quad 1\leq m\leq 2^l-1, \ \  1\leq l \leq n-1.
\end{align}
Thus $f(\cdot)$ is not necessarily an increasing function. $\blacksquare$
\end{fact}

\begin{fact}\label{fact:weight}
For a given polar code of length $N$ formed by the kernel $F$, and for any $i\in \mathcal{A}$, we have $f(i)=wt(\textbf{r}_i)$. In other word, the size of the leaf set for any stopping tree is in fact equal to the weight of the corresponding row in the generator matrix. Particularly, the leaf set of the stopping tree for any input bit represents the locations of 1's in the corresponding row of the matrix $F^{\otimes n}$. $\blacksquare$
\end{fact}

Now, let us consider variable-node stopping sets for $J\subseteq \mathcal{A}$. The following theorem is proved for $MVSS(J)$ in the Appendix. The proof uses facts \ref{fact:stopends}, \ref{fact:stopinduce}, and \ref{fact:f}.

\begin{thm}\label{th:MVSS}\label{th:minVSS}
Given any set $J\subseteq \mathcal{A}$ of information bits in a polar code of length $N=2^n$, we have $|MVSS(J)|\geq \min_{j\in J} f(j)$. $\blacksquare$
\end{thm}

Theorem \ref{th:minVSS} sets a lower bound on the size of the $MVSS$ for a subset $J$ of information bits. It also implies that the size of the minimum VSS for a polar code is at least equal to $\min_{i\in \mathcal{A}} f(i)$. However, we already know that the leaf set of the stopping tree for any node $i\in \mathcal{A}$ is a VSS of size $f(i)$. This leads us to the following corollary.

\begin{cor}\label{cor:minVSS}
For a polar code with information bit index $\mathcal{A}$, the size of a minimum variable-node stopping set is equal to $\min_{i\in \mathcal{A}} f(i)$, i.e. the size of the leaf set for the minimum-leaf stopping tree. $\blacksquare$
\end{cor}

Corollary  \ref{cor:minVSS} implies that in order to find the size of the minimum VSS, we need to find the information bit with minimum leaf stopping tree among all the information bits.

\subsection{Size Distribution of Stopping Trees and Their Leaf Sets}\label{sec:sizedist}
We provide a method for finding the size distribution of stopping trees and their leaf sets.
First, note that the recursive construction of the factor graph dictates a relationship between the size of  stopping trees in $T_{n+1}$ and $T_n$.

\begin{fact}\label{fact:sizedist}
Let $\textbf{A}_n$ and $\textbf{B}_n$ be two vectors of length $2^n$ showing, respectively, the size of stopping trees and their leaf sets for all input bits in $T_n$. That is, $\textbf{A}_n=[|ST(1)|\ |ST(2)|\ ...\ |ST(2^n)|]$ and $\textbf{B}_n=[f(1)\ f(2)\ ...\ f(2^n)]$. We then have
\begin{align}\label{eq:sizedist}
\nonumber \textbf{A}_{n+1} &= [\textbf{A}_n \  2\textbf{A}_n]+\textbf{1}_{n+1}\\
\textbf{B}_{n+1}&=[\textbf{B}_n \  2\textbf{B}_n],
\end{align}
where $\textbf{1}_{n+1}$ is the all-ones vector of length $2^{n+1}$. $\blacksquare$
\end{fact}

These two recursive equations can be solved with complexity $O(N)$ to find the desired size distributions for a code of length $N$. Note that Fact \ref{fact:f} can be also concluded from Fact \ref{fact:sizedist}.
Furthermore, Fact \ref{fact:weight} can be used to find the size of leaf set for a specific stopping tree within time $O(N)$.

\subsection{Stopping Distance for Polar Codes}
Fact \ref{fact:sizedist} gives the stopping distance for a finite-length polar code, when the set of information bits is known. However, it is not always easy to choose the optimal information set, particularly with large code-lengths.
In order to approach this problem, we first show that a slight modification in the set of information bits may actually result in a larger stopping distance without a significant impact on the BER performance.

\begin{thm}\label{th:vanish}
In the factor graph of a polar code of length $N$, the number of input bits $v(i,1)$ for which $f(i)<N^\epsilon,\ 0<\epsilon<\frac{1}{2}$ is less than $N^{H(\epsilon)}$. $\blacksquare$
\end{thm}

\begin{figure}[t]
\centering
{\includegraphics[width =4 in , height=3 in]{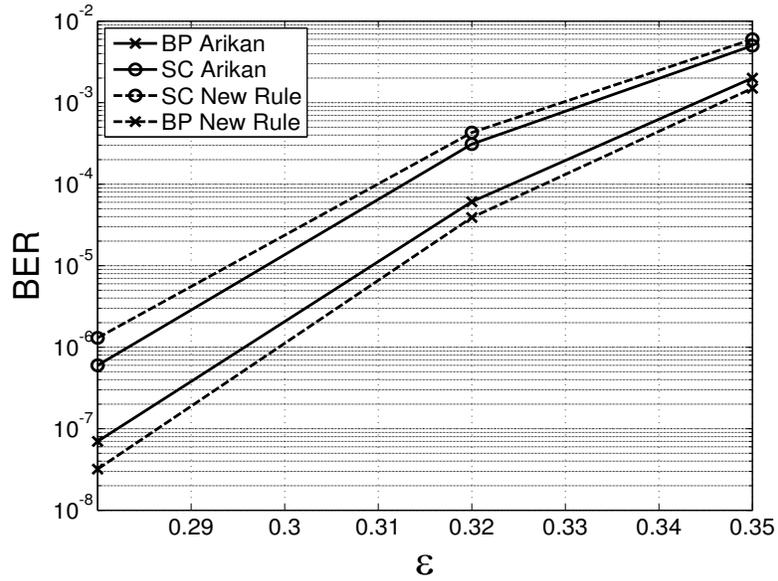}}
\caption{BER comparison for different methods of choosing information bits under BP and SC decoding. Code-rate and code-length are 1/2 and $2^{13}$, respectively.}
\label{fig:newrule}
\end{figure}

The above theorem implies that, for any $0<\epsilon<1/2$, we can always replace $N^{H(\epsilon)}$ information bits by some frozen bits for which the stopping tree has a leaf set larger than $N^\epsilon$.
It is easy to show that such a replacement does not effectively change the overall BER under BP, asymptotically.
When $N\rightarrow \infty$ and $\epsilon<1/2$ ,  $N^{H(\epsilon)}$  will be vanishing with $N$.
In a sparse factor graph, such as the one in polar codes, erroneous decoding of a small set of information bits affects only a few number (vanishing with $N$ as $N\rightarrow \infty$) of other information bits. Therefore, given a finite number of iterations, BER will not change asymptotically.
Accordingly, We can expect such a modification to have little impact on the BER performance in the finite regime, while resulting in a better error floor performance.
Fig. \ref{fig:newrule} is used to demonstrate this case. The BER is depicted for Arikan's rule and its modified version introduced above (we call it \emph{new rule}) applied to a code of length $2^{13}$ and rate 1/2. We replaced information bits with leaf sets smaller than $2^8$, by frozen bits with minimum Bhattacharyya parameter who also had a leaf set larger than $2^8$. As it can be seen, when SC decoding is used, the new rule performs slightly worse than the Arikan's rule. However, under BP decoding, it does slightly better than Arikan's rule. While the figure only shows the BER performance in the waterfall region, We conjecture that this rule results in a superior error floor performance of the new rule due to its larger stopping distance. It is also noteworthy that if we use the new rule to pick all the information bits, i.e. if we only pick input bits with largest leaf sets as information bits, then the resulting code will be a Reed-Muller code for which BP performance is worse than polar codes \cite{arikan08}. Therefore, we only considered a limited use of the new rule. This apparently helps to preserve some of the good characteristics of polar codes while increasing the stopping distance.
We also like to mention two points regarding the stopping distance.

\subsubsection{Asymptotic Case}
Theorem \ref{th:vanish} asserts that given any capacity-achieving polar code and any $\sigma>0$, we can always construct another capacity-achieving code with a stopping distance $N^{1/2-\sigma}$, by replacing some information bits by some frozen bits with larger $f(.)$. The following theorem gives the stopping distance for polar codes in the asymptotic case. Note that this only holds asymptotically and the analysis is different for finite-length codes, as we explained above.

\begin{thm}\label{th:stopasymp}
The stopping distance for a polar code of length $N$ grows as $\Omega(N^{1/2})$. $\blacksquare$
\end{thm}

\subsubsection{Minimum Distance vs. Stopping Distance}
The following theorem states the relation between the stopping distance and minimum distance of polar codes.
\begin{thm}\label{th:stopdmin}
The stopping distance of a polar code defined on a normal realization graph such as the one in Fig. \ref{fig:stopset1}, is equal to the minimum distance of the code, $d_{min}$. $\blacksquare$
\end{thm}


According to Theorem \ref{th:stopdmin}, the number of code-bits in the minimum VSS grows as fast as the minimum distance.
It is noteworthy that for linear block codes, $d_{min}$ (i.e. the minimum Hamming weight among all codewords) puts an upper bound on the stopping distance \cite{Di06,Orlitsky05,Orlitsky02}. This is because if all the ones in the received vector are erased, then it is impossible for the decoder to find out if an all-zero codeword has been sent or another codeword.
For a code, it is a desirable property to have a stopping distance equal to its minimum distance.
Therefore, Theorem \ref{th:stopdmin} can be interpreted as a positive result, particularly compared to the capacity-approaching LDPC codes for which both the stopping and minimum distances are fairly small in comparison to the block length \cite{Di06,Orlitsky05,Orlitsky02}.

\section{Error Floor Performance of Polar Codes}\label{sec:errorfloor}
A large stopping distance is desirable in order to improve the error floor performance of a code over the BEC. After exploring the stopping sets of polar codes in the pervious section, here we focus on ``girth" of polar codes as another important factor in error floor performance. Afterward, we examine the error floor performance of polar codes over the BEC and binary Gaussian channel via simulations.
\subsection{Girth of Polar Codes}
\begin{figure}[t]
\centering
{\includegraphics[width =5 in , height=3.5 in]{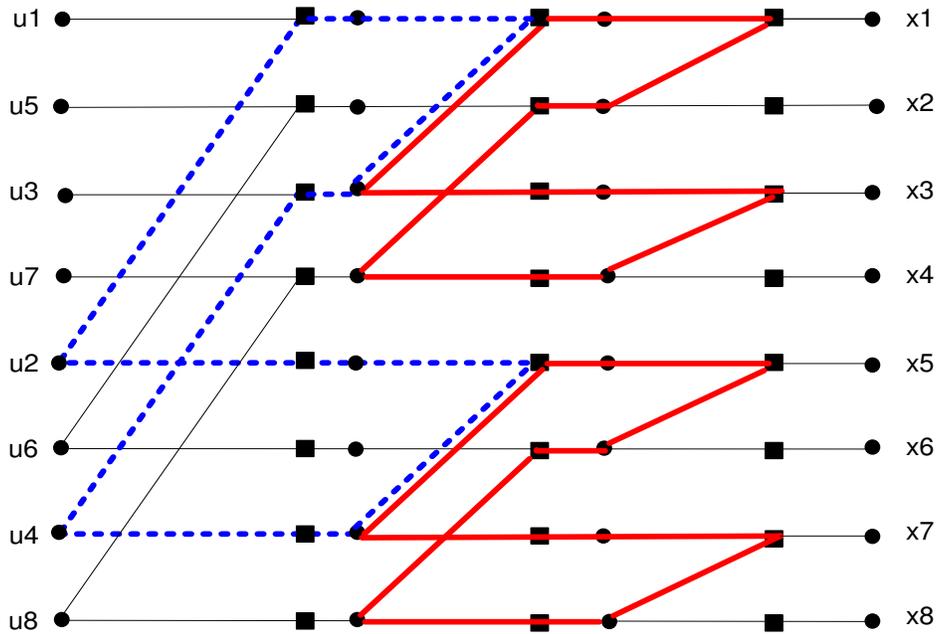}}
\caption{Different types of cycles in the factor graph of polar codes for $N=8$. Thick solid and dashed lines show the first and second types of cycles, respectively.}
\label{fig:cycle1}
\end{figure}

The \emph{girth} of a graph is the length of shortest cycle contained in the graph.
cycles in the Tanner graph prevent the sum-product (BP) algorithm from converging \cite{KschischangFactor01}. Furthermore, cycles, especially short ones, degrade the performance of the decoder, because they affect the independence of the extrinsic information exchanged in the iterative decoding.
When decoded by belief propagation, the external information at every variable node remains uncorrelated until the iteration number reaches half the girth. Hence, we are often interested in constructing large girth codes that can achieve high performance under BP decoding \cite{Gholami12, Bocharova12, Huang10Girth}.
As it can be seen in the factor graph shown in Fig. \ref{fig:cycle1}, there exist two types of cycles: first, the cycles including nodes only from one of the top or bottom part of the graph (shown by thick solid lines), and second, the cycles including nodes from both top and bottom parts of our symmetric graph (shown by thick dashed lines). The first type of cycles have the same shape in both upper and lower halves of the graph. The interesting fact about the cycles is that because the graph for a code of length $2^m$ is contained in the graph of a code of length $2^{m+1}$, all the cycles of the shorter code are also present in the graph of the longer code.
The shortest cycle appears in the graph of a length-4 polar code, as it is shown in Fig \ref{fig:cycle1}. It is a cycle of size 12, including 6 variable nodes and 6 check nodes.
The shortest cycle of the second type appears first in the graph of a length-8 polar code, and have a size of 12 (dotted lines in Fig. \ref{fig:cycle1}). Thus, based on the above, the girth of a polar code is 12.


\subsection{Simulation Results for Error Floor}
\begin{figure}[t]
\begin{center}
\subfigure[BER for BP and SC decoding over BEC. The code-length and code-rate are $2^{15}$ and  $1/2$, respectively. The $99\%$ confidence interval is shown for the two lowest BER's.]{\includegraphics[width =5.5 in , height=3.6 in]{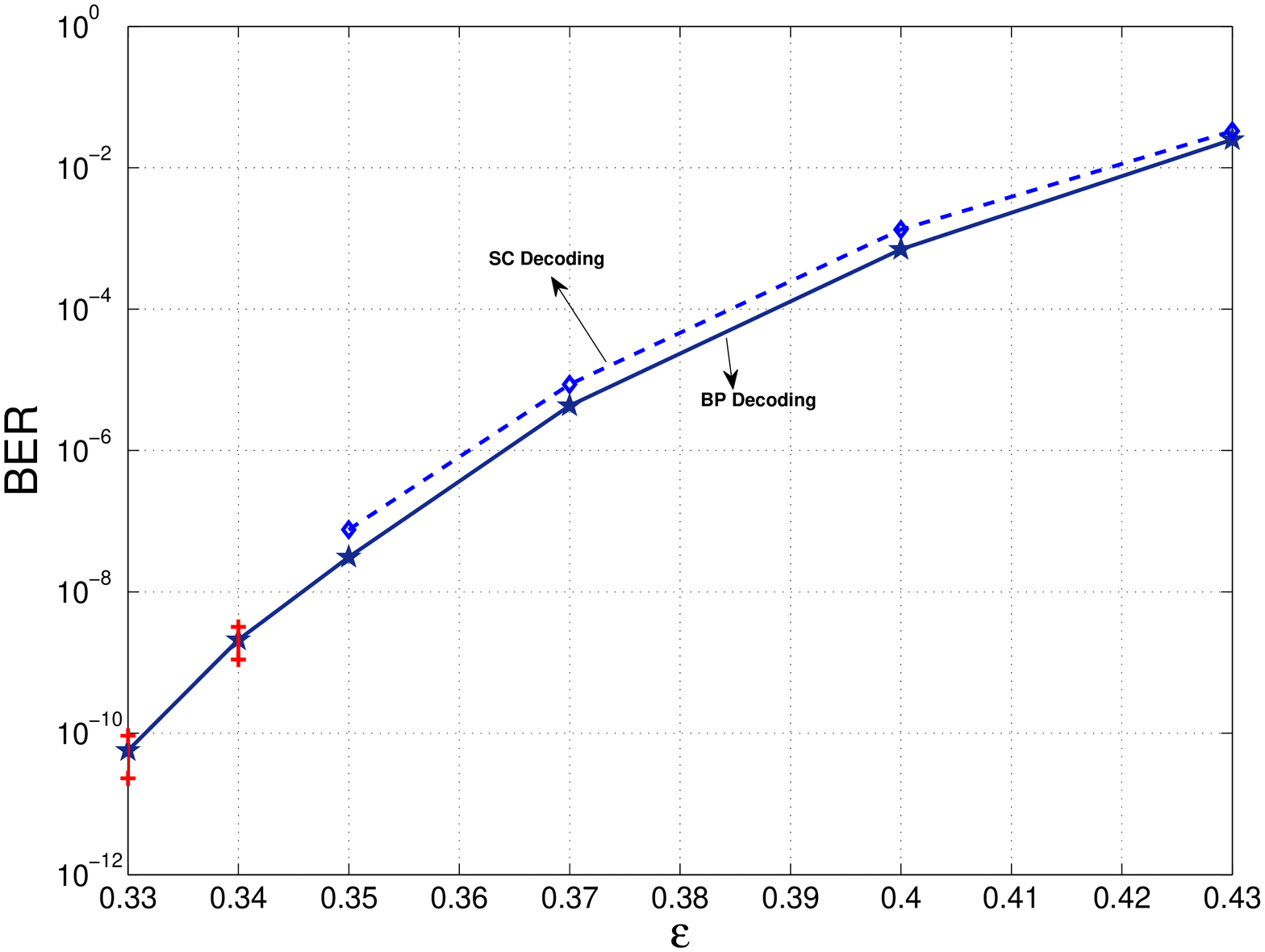} \label{fig:bererasure}}
\subfigure[BER for BP and SC decoding over Gaussian channel. The code-length and code-rate are $2^{13}$ and  $1/2$, respectively.]{\includegraphics[width =5 in , height=3.6 in]{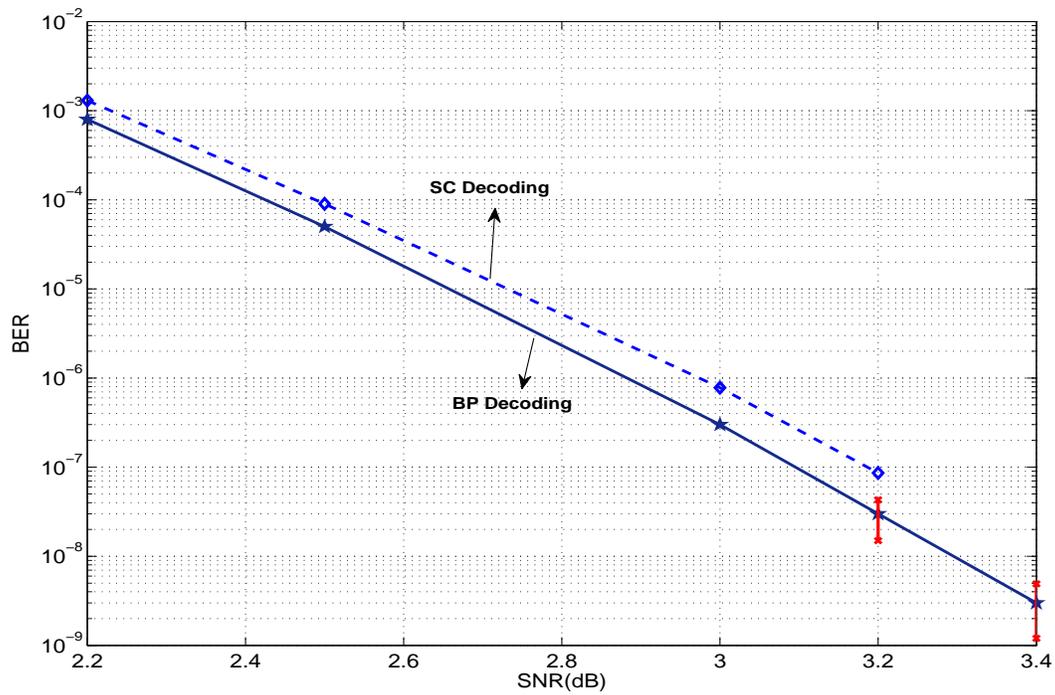} \label{fig:bergauss}}
\end{center}
\label{fig:bers}
\caption{BER performance of polar codes over the binary erasure and Gaussian channels. The $99\%$ confidence interval is shown for the two lowest BER's.}
\end{figure}

We performed simulations to examine the effect of the relatively large stopping distance and girth of the polar codes' factor graph on the error correction performance of these codes. Fig. \ref{fig:bererasure} shows the simulation results for a code of length $2^{15}$ and rate 1/2 over the BEC. As it can be seen, no sign of error floor is apparent. This is consistent with the relatively large stopping distance of polar codes.
We indicated the $99\%$ confidence interval for low BERs on the curve to show the precision of the simulation.
Fig. \ref{fig:bergauss} also shows the simulation results for a rate $\frac{1}{2}$ polar code of length $2^{13}$ over a binary-input Gaussian channel subjected to additive white Gaussian noise with zero mean and variance $\sigma^2$. The figure shows no sign of error floor down to the BERs of $10^{-9}$.

Regarding the error floor, we should mention here a prior work by Mori and Tanaka \cite{mori09}, which gives theoretical upper and lower bounds on \emph{block error probability}, for SC decoding of polar codes over the BEC. According to these bounds, no error floor is expected for block error probability.
Note also that for the BEC, BP decoding is strictly better than SC decoding \cite{hussami09}. Thus, if SC decoding shows no error floor problems, so does BP decoding. For large block lengths, however, a stopping distance of $\Omega(\sqrt{N})$ (as it was shown in Theorem \ref{th:stopasymp}) implies a good error floor performance for polar codes over the BEC.


\section{A Potential Application for Polar Codes}\label{sec:concat}
Polar codes show a set of good characteristics that are needed in many real-world communication systems. Among these properties are good error floor performance, being capacity-achieving, and a low encoding and decoding complexity. In this section, we take advantage of these properties to design a polar code-based scheme as a solution to a practical problem. An \emph{Optical Transport Network (OTN)} is a set of optical network elements connected by optical fiber links, able to transport client signals at data rates as high as 100 Gbit/s and beyond. These networks are standardized under ITU-T Recommendation G.709, and stand for an important part of the high data-rate transmission systems such as Gigabit Ethernet and the intercontinental communication network.
A minimum BER of at least $10^{-13}$ is generally required in such systems \cite{ningde08,Mizuochi09}. Because of very high-rate data transmission, OTNs need to employ a low complexity coding scheme to keep the delay in a low level. Furthermore, these systems generally use a long frame for data transmission, which allows using large code-lengths.

\begin{figure}[t]
\centering
{\includegraphics[width=4.5 in , height=1.8 in]{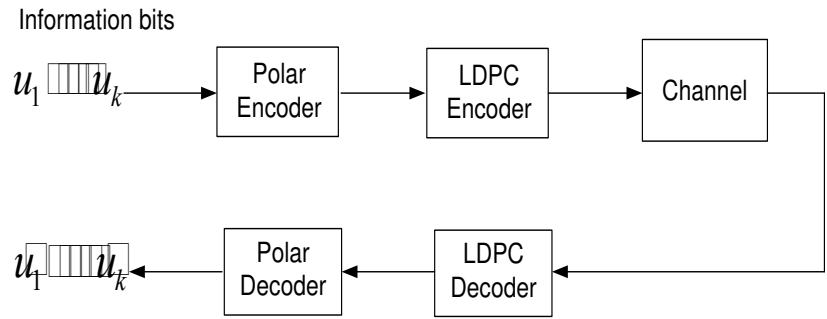}}
\caption{Block diagram of the proposed concatenated system of polar and LDPC codes.}
\label{fig:concat}
\end{figure}

We propose concatenated polar-LDPC codes to be used in OTNs. Our proposed scheme is formed of a Polar code as the outer code, and a LDPC code as the inner code. Fig. \ref{fig:concat} shows the block diagram of this scheme.
We consider long powerful LDPC codes as the inner code with rates close to the channel capacity.
LDPC codes with good waterfall characteristics are known to mostly suffer from the error floor problem.
However, the polar code plays a dominant role in the error floor region of the LDPC code.
Based on the analysis provided in previous sections, the combination of polar and LDPC codes is expected to form a powerful concatenated scheme with a BER performance close to the capacity for a broad range of the channel parameter.
We consider a binary polar code concatenated with a binary LDPC code. This is different from the traditional concatenated schemes \cite{lincostello83} in which a non-binary code is usually used as the outer code.

OTU4 is the standard designed to transport a 100 Gigabit Ethernet signal.
The FEC (Forward Error Correction) in the standard OTU4 employs a block interleaving of 16 words of the (255, 239, 17) Reed-Solomon codes, resulting in an overall overhead of 7\%. This scheme guarantees an error floor-free performance at least down to BERs of $10^{-15}$, and provides a coding gain of 5.8 dB at a BER of $10^{-13}$. Since the approval of this standard (February 2001), several concatenated coding schemes have been proposed in the literature and some as patents, targeting to improve the performance of this standard. In most cases, these schemes propose a concatenation of two of Reed-Solomon, LDPC, and BCH codes \cite{ningde08,wu10,Mizuochi09,Griesser}.
Here, for the first time, we consider polar-LDPC concatenation for the OTU4 setting.

\subsection{Encoder}
\begin{figure}[t]
\centering
{\includegraphics[width =4.5 in , height=3.5 in]{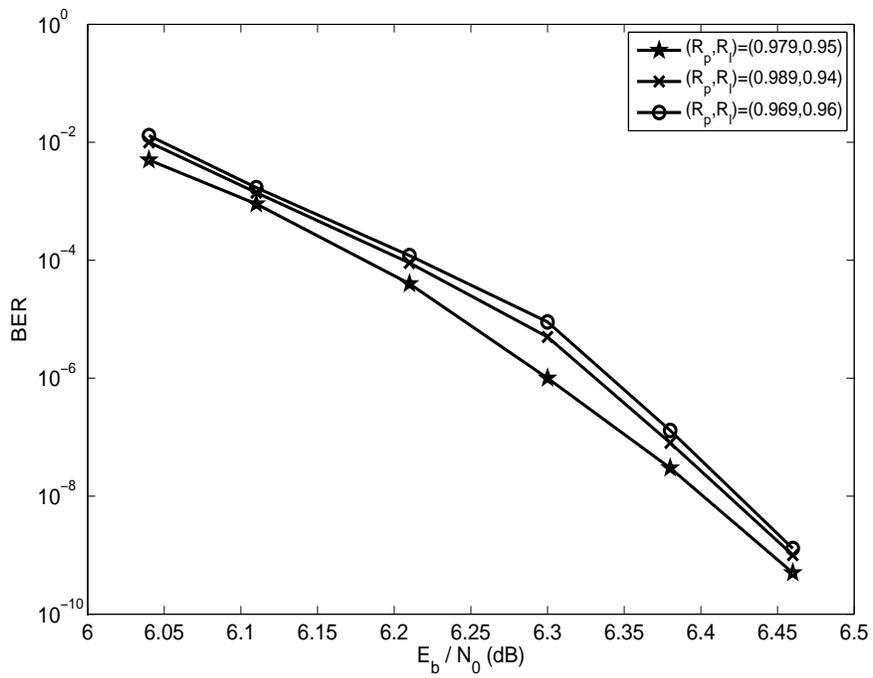}}
\caption{BER performance comparison for different rate combinations in a polar-LDPC concatenated scheme.}
\label{fig:ratecomb}
\end{figure}

In order to satisfy the overhead of 7\%, we adopt an effective code rate of 0.93. That is, if we denote the code-rates for the polar and LDPC codes by $R_p$ and $R_l$ respectively, then $R_{eff}=R_p\times R_l$ needs to be 0.93.
The first problem is to find the optimal code-rate combination for the two codes to achieve the best BER performance. While this is an interesting analytical problem, it might be a difficult problem to solve. Therefore, we find the best rate combination for our application empirically. First, note that both $R_p$ and $R_l$ are greater than 0.93. We are also aware of the relatively poor error rate performance of finite-length polar codes compared to LDPC codes. Therefore, in order to minimize the rate loss, we choose $R_l$ close to the $R_{eff}$. As a result, $R_p$ would be close to 1. The values of $R_l$ and $R_p$ can be found empirically. Fig. \ref{fig:ratecomb} shows the BER performance of three different rate couples, as a sample of all the rate couples we simulated. Code-length for the polar code is fixed to $2^{15}=32768$ for all the rate couples. Showing a rate couple by $(R_p, R_l)$, these three rate couples are (0.989, 0.94), (0.979, 0.95), (0.969, 0.96). We picked (0.979, 0.95) for the rest of our simulations in this paper as it shows a better performance in the low-error-rate region. Fixing the code-length $2^{15}=32768$ for the polar code and fixing the rates to (0.979, 0.95), the LDPC code-length would be 34493. We used the following optimal degree distribution pair which has a threshold value of $0.47$ for the binary AWGN channel under BP  \cite{sigpromu}:
\begin{align}
\nonumber \lambda(x)&=0.156935\ x+ 0.138295\ x^2+ 0.325131\ x^3+ 0.168818\ x^{11} + 0.210821\ x^{12},\\
\nonumber \rho(x)&=0.039239\ x^{34}+ 0.144375\ x^{35} + 0.302308\ x^{70}+ 0.514078\ x^{71}.
\end{align}

An interesting question here is how to design the polar code in this concatenated scheme, while the channel seen by the polar code is not an AWGN channel anymore.
It is well known, that when the iterative BP decoder fails, the residual erroneous bits after decoding are organized in graphical structures (e.g. stopping sets on BEC or trapping sets for other types of channels). In order to find the distribution of such patterns, one method is to prepare a histogram of these (post-decoding) error patterns. However, here we simply assume that the error patterns are distributed randomly (equally likely) at the output of the LDPC decoder, hence assuming the channel seen by the polar code as an AWGN channel with capacity 0.979. We then designed our polar code for this channel.
The problem of designing optimal polar codes for this concatenated scheme remains as an interesting problem for further research.

\subsection{Decoder}
At the decoder side, we perform belief propagation decoding with soft-decision for both the polar and LDPC codes. Upon finishing its decoding, the LDPC decoder will pass its output vector of LLRs to the polar decoder. Polar decoder then treats this vector as the input for its belief propagation process.

\subsection{Simulation Results}
\begin{figure}[t]
\centering
{\includegraphics[width =6 in , height=4.8 in]{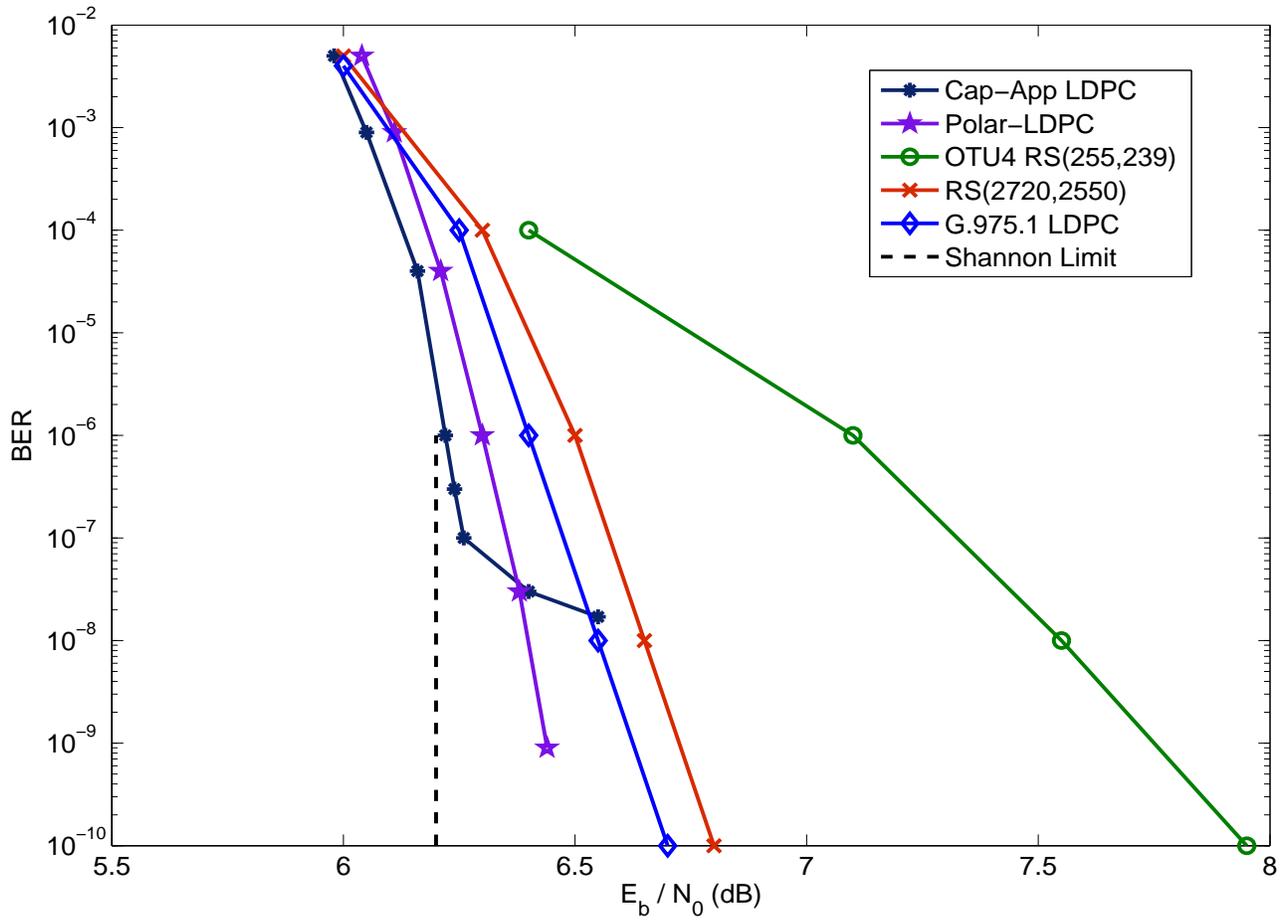}}
\caption{BER performance for different concatenated schemes.}
\label{fig:casc1}
\end{figure}

Fig. \ref{fig:casc1} depicts the BER performance for the concatenated scheme explained above, when using the LDPC code above. For the channel, we assumed a binary symmetric Gaussian channel as it is used by \cite{ningde08,wu10,Mizuochi09,Griesser}.
Along with the concatenated scheme, we have shown the performance of the LDPC code when used alone with an effective rate of 0.93, which is equal to the effective rate of the concatenated scheme.
As it can be seen, the concatenated scheme follows the performance of LDPC code in the waterfall region closely. Since both polar and LDPC codes here are capacity-approaching (capacity-achieving in case of polar codes), this technique does not suffer from rate-loss theoretically. Therefore, by increasing the code-length we expect the curve for polar-LDPC scheme to close the gap to capacity. The curve also shows no sign of error floor down to BERs of $10^{-10}$, as opposed to the curve for LDPC code which shows error floor at around $10^{-8}$.
What actually happens in a polar-LDPC concatenation is that the two codes are orchestrated to cover for each other's shortcomings: LDPC plays the dominant role in its waterfall region, while polar code is dominant in the error floor region of the LDPC code.

We should also mention that a soft BP decoder is used with a 9 bit quantization (512 values) of the LLRs. We are also limiting the LLR values to the range of (-20, 20). The maximum number of iterations used in our simulations is 60; however, we counted the average number of iterations (let us call it the ANI) for LDPC and polar-LDPC schemes in order to get some ideas about their decoding latency. At a BER of $10^{-6}$, the ANI for the capacity-approaching LDPC code when used alone was 11.3. On the other hand, the ANI for the LDPC and polar codes used in the polar-LDPC scheme was 13.1 and 16.7, respectively. It should be noted that the BP-Polar iterations are heavier than the iterations for LDPC due to the $N\log N$ time of each iteration in BP-Polar in comparison to the linear time of each iteration in BP-LDPC. In our simulations for the lower points in the curves, we kept sending blocks until we encounter 100 erroneous blocks. For example, for polar-LDPC curve at 6.4 dB (the lowest BER), we ended up simulating over 300 million blocks. This particular point took us the longest amongst all the simulated points. The lowest point in the cap-app LDPC curve was obtained by simulating about 30 million blocks.

In order to see the significant potential of polar codes for concatenated schemes, we compared the BER performance of the polar-LDPC approach against some of the existing coding techniques for OTNs, including the G.709 standard explained earlier in the paper.
We also included two ``super FECs" proposed in ITU-T standard G.975.1 for high bit-rate DWDM (Dense Wavelength Division Multiplexing) submarine systems \cite{G9751}.
These schemes share some features, specifically the rate, block-length, and low decoding latency, with G.709, while achieving a much better performance.
All the schemes use a code rate of 0.93. Furthermore, all of them are using codes of length around $2^{15}$. We borrowed the BER curves of these schemes from \cite{G9751}.

As it is shown, an improvement of 1.3 dB at BER of $10^{-8}$ is achieved  by polar-LDPC over the RS(255,239) of G.709 standard. Another scheme is an RS(2720,2550) with 12-bit symbols that has a block-length of 32640 bits. It has been shown to achieve a significant coding gain and to have superior burst correction capabilities \cite{G9751}. As it is shown, polar-LDPC concatenation achieves an improvement of 0.25 dB over this scheme. Presented in the figure is also the performance of a systematic binary LDPC code of length 32640, with 30592 information-carrying bits \cite{G9751}. This LDPC code is suitable for implementation in current chip technologies for 10G and 40G optical systems offering low latency and feasibility of low power consumption in case of 40G implementation showing a significantly higher coding gain than the standardized RS code in G.709. As it can be seen, polar-LDPC shows an edge of 0.15 dB over this LDPC scheme.
The decoding complexity for LDPC and RS codes is $O(N)$ and $O(N^2)$, respectively, while the polar-LDPC scheme has a complexity of $O(N \log{N})$ which is closer to the LDPC code.


\section{Conclusion}\label{sec:conclusion}
As a first step in a practical approach to polar codes, we studied the BER performance of finite-length polar codes under belief propagation decoding. We analyzed the structure of stopping sets in the factor graph of polar codes as one of the main contributors to the decoding failure and error floor over the BEC. The size of the minimum stopping set and the girth of the factor graph have been found for polar codes. We then investigated the error floor performance of polar codes through simulations where no sign of error floor was observed down to BERs of $10^{-10}$.
Motivated by good error floor performance, we proposed using polar codes in combination with other coding schemes. We particularly studied the polar-LDPC concatenation to be used in OTNs as a potential real-world application. Comparing the performance for our proposed scheme to some of the existing coding schemes for OTNs, we showed that polar-LDPC concatenation can achieve a significantly better performance.

{
\bibliographystyle{ieeetr}
\linespread{1.3}
\bibliography{hldpcr,hldpcr1}}

\appendix

\emph{Proof of Fact \ref{fact:stopends}:}
First, note that we only have degree 2 and 3 check nodes in the graph. In every Z-shaped part there are two check nodes, one at the top and one at the bottom. The top check node is always of degree 3 and the bottom one is always of degree 2. When a check node is a neighbor of a variable node or a set of variable nodes, we say that the check (variable) node is \emph{adjacent to} that variable (check) node or the set of variable (check) nodes.
We show that if a GSS is adjacent to either one of these check nodes in the $i$th column, then it must involve check nodes and variable nodes from both $(i-1)$th and $(i+1)$th columns. Therefore, any GSS includes variable nodes from all columns of the graph, including information bits and code-bits.

We consider two cases. Since each neighboring check node of a GSS needs to be connected to at least two variable nodes in the set, if the bottom check node is adjacent to the GSS, then both of its neighboring variable nodes must be in the set. Since all the check nodes connected to a variable node in the GSS are also adjacent to the set, this means that some of the check nodes in the $(i-1)$th and $(i+1)$th columns are also adjacent to the set. In the second case, if the upper check node (of degree 3) is adjacent the GSS, then its neighbors in the GSS are either a variable node at its right and one at its left, or two variable nodes at its left, one at the top and one at the bottom of the Z.
In the former case, the GSS clearly includes nodes from the $(i-1)$th and $(i+1)$th columns. In the latter case, the bottom variable node has the bottom check node as its neighbor in the GSS, leading to the same situation we discussed above. $\blacksquare$

\begin{figure}[t]
\centering
{\includegraphics[width =4 in , height=3.2 in]{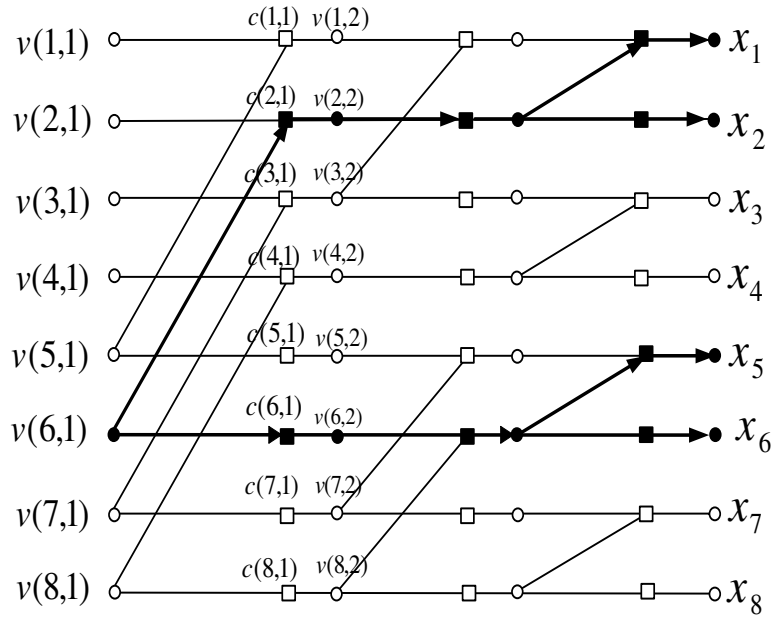}}
\caption{The stopping tree for $v(6,1)$ is shown  with black variable and check nodes.}
\label{fig:stopset2}
\end{figure}

\vspace{.1 in}
\emph{Proof of Fact \ref{fact:uniqstoptree}:}
Suppose an information bit $i$ has two non-overlapping stopping trees, $ST$ and $ST^\prime$.
Also, suppose $ST$ has a form like the stopping tree shown in Fig. \ref{fig:stopset2}. That is only one variable node from each Z can participate in $ST$. Also, Note that a check (variable) node in the graph is adjacent to only one variable (check) node on the right (left). Thus, if a check node is adjacent to $ST$, it is adjacent to exactly one variable node on the left and one on the right.

Now assume that the difference between $ST$ and $ST^\prime$ starts at the $j$th column. $j\neq1$ Since, by definition, a stopping tree can include only one information bit; hence, $v(i,1)$ is the only variable node of column 1 participating in $ST$ and $ST^\prime$.
Suppose there exists a variable node $v(k^\prime,j)\in ST^\prime ,j\neq1,$ which is not part of $ST$. $v(k^\prime,j)$ is adjacent to $c(k^\prime,j-1)$ from left. However, $c(k^\prime,j-1)$ can not be adjacent to $ST$, otherwise we would have $v(k^\prime,j)\in ST$ because of what we mentioned above. But $c(k^\prime,j-1)$ must be adjacent to at least one variable node in $ST^\prime$ form the left since it needs to be adjacent to at least two variable nodes in $ST^\prime$ (definition of a stopping set). Therefore, $c(k^\prime,j-1)$ is adjacent to at least one variable node in $ST^\prime$ in the $(j-1)$th column, which is not part of $ST$. This is contradiction since we assumed $ST$ and $ST^\prime$ start to differ at the $j$th column. $\blacksquare$

\vspace{.1 in}
\emph{Proof of Fact \ref{fact:stopinduce}:}
Fact \ref{fact:stopends} implies that any GSS in $T_{n+1}$ includes at least one information bit. Consider such a GSS.
According to Fact \ref{fact:stopends}, this GSS includes a set of variable nodes in $T_{n+1}^{U}$ and/or $T_{n+1}^{L}$. Let us denote these sets by $S^{U}$ and $S^{L}$, respectively.
Now, it is easy to see that the variable and check nodes in $S^{U}$ and $S^{L}$, if non-empty, still satisfy the conditions of a GSS.
This is because $v(1,1), v(2,1), ..., v(N,1)$ are connected to the rest of the graph only through $c(1,1), c(2,1), ..., c(N,1)$. Therefore, for any GSS in $T_{n+1}$, the induced non-empty subsets in $T_{n+1}^{U}$ and $T_{n+1}^{L}$ also form a GSS for these subgraphs. $\blacksquare$

\vspace{.1 in}
\emph{Proof of Fact \ref{fact:f}:}
This fact can be concluded directly by looking at the recursive structure of the factor graph. $\blacksquare$

\vspace{.1 in}
\emph{Proof of Fact \ref{fact:weight}:}
This is true because based on Arikan's paper, the encoding graph of polar codes is obtained from the matrix $F^{\otimes n}$. In fact, this graph is a representation of the recursive algebraic operations in this Kronecker product. $\blacksquare$

\vspace{.1 in}

\begin{figure}[t]
\centering
{\includegraphics[width =4 in , height=3.5 in]{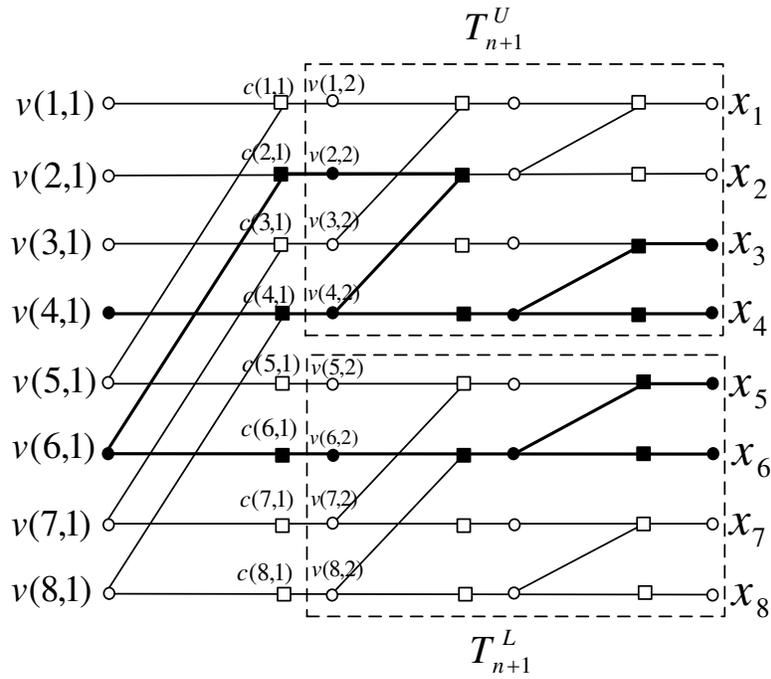}}
\caption{Normal realization of the encoding graph for $N=8$. An example of a GSS is shown with black variable and check nodes.}
\label{fig:stopset1}
\end{figure}

\begin{figure}[t]
\begin{center}
\subfigure[Case 1 in Theorem \ref{th:minVSS}.]{\includegraphics[width =4 in , height=3.4 in]{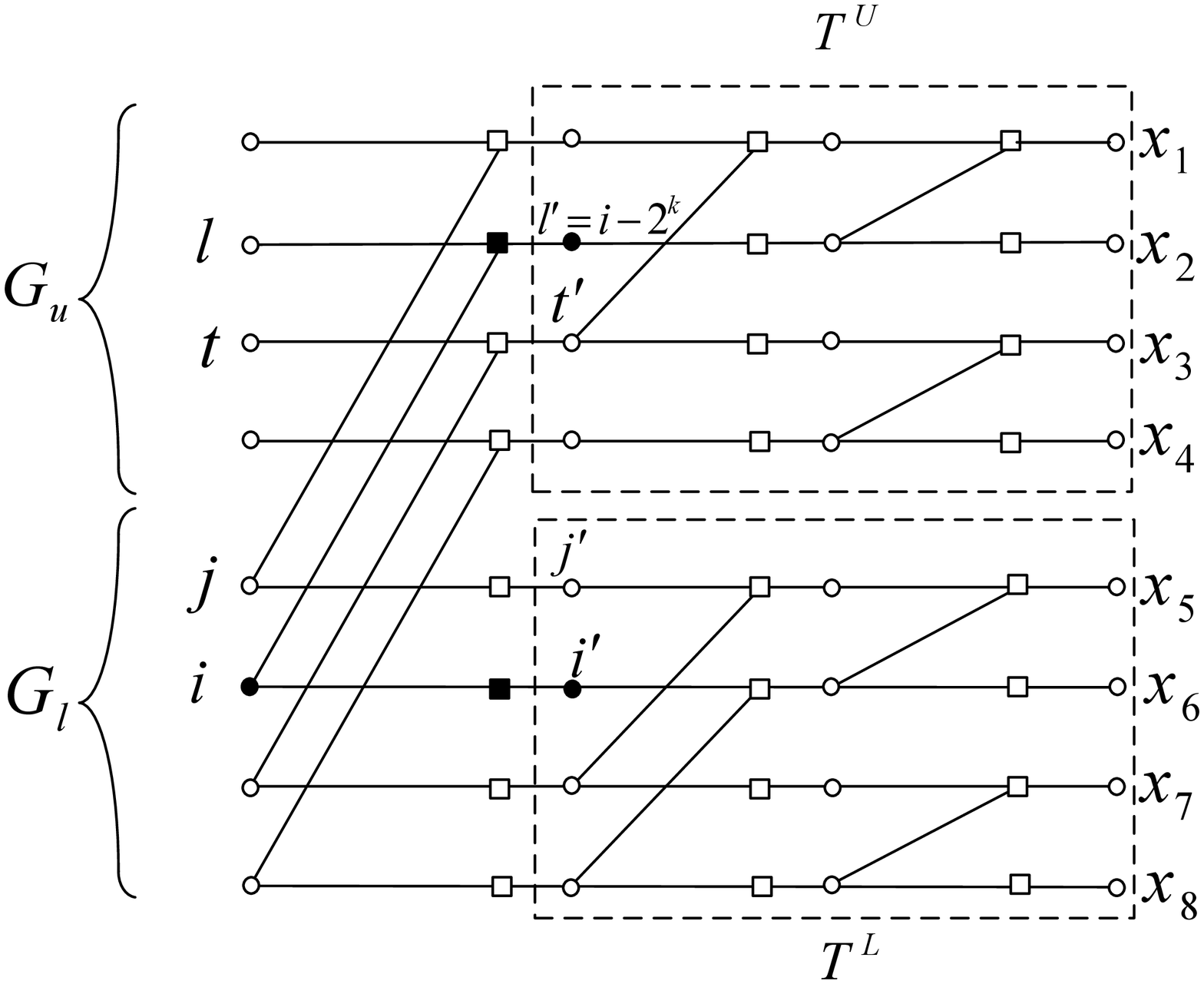} \label{fig:stopset3}}
\subfigure[Case 2 in Theorem \ref{th:minVSS}.]{\includegraphics[width =4 in , height=3.4 in]{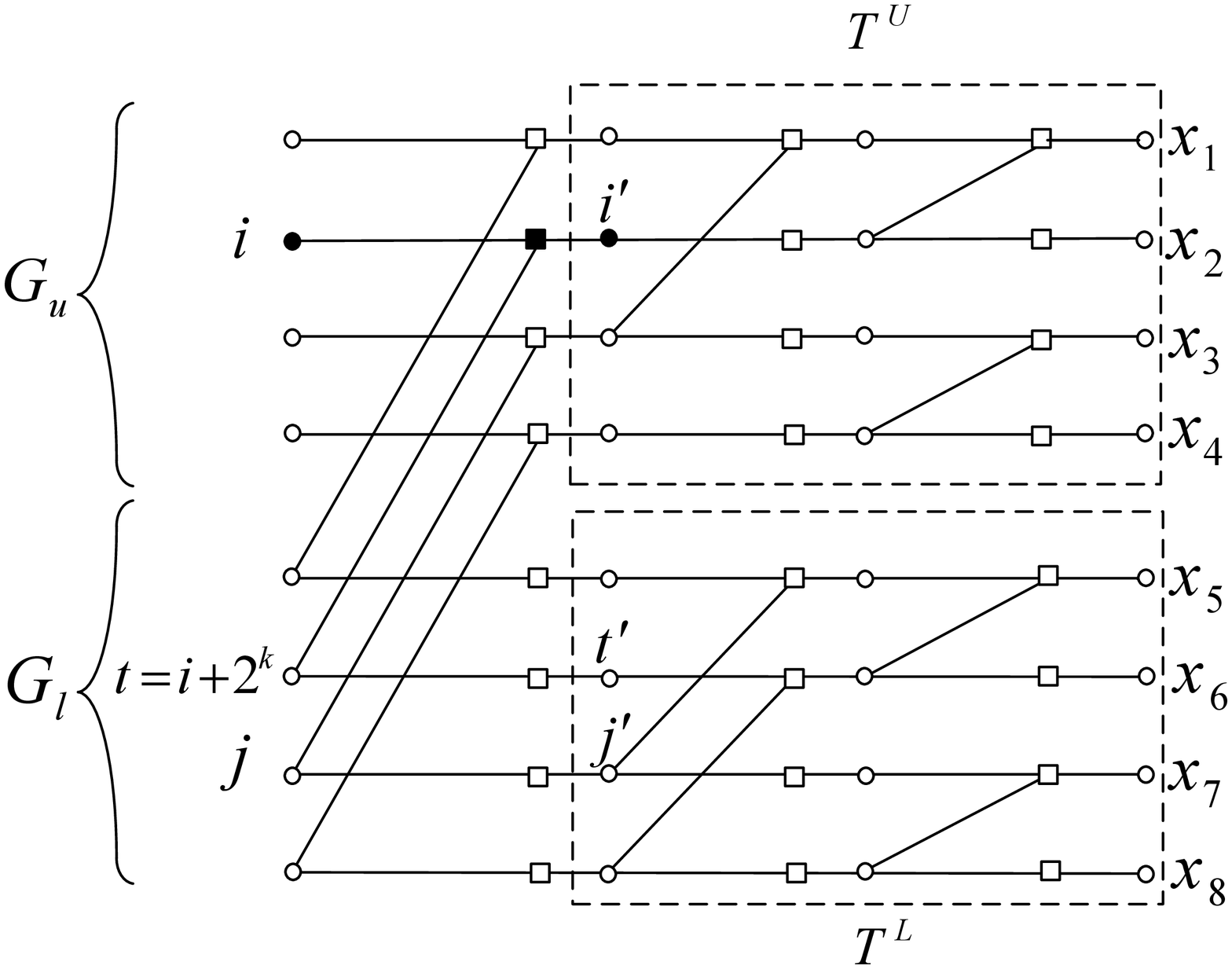} \label{fig:stopset4}}
\end{center}
\label{fig:MSSVproof}
\caption{Figure is used to visualize different cases considered in the proof of Theorem \ref{th:minVSS}.}
\end{figure}

\emph{Proof of Theorem \ref{th:MVSS}:}
We prove the theorem by induction on $n$ where $N=2^n$ is the code-length. For $n=1$ ($N=2$), there are only two information bits, $v(1,1)$ and $v(2,1)$. It is trivial to check the correctness of the theorem in this case. Now suppose the hypothesis holds for a polar code of length $2^k$. We prove that it also holds for a code of length $2^{k+1}$. Consider a set $J$ and let $MIB(J)=i$.
In the case that there exist more than one MIB in $J$, without loss of generality, we pick the one with the largest index as the $MIB(J)$. That is, we pick the one which occupies the lowest place in the graph among the MIBs of $J$.
Let $VSS^*$ be a minimum VSS for $J$, and let $GSS^*$ be the corresponding GSS for $VSS^*$.
We also denote the upper and lower halves of the factor graph by $G_U$ and $G_L$, as it is shown in Fig. \ref{fig:stopset3}. Note that $G_U$ and $G_L$ are identical in shape, and each of them includes half of the variable and check nodes in the factor graph.
Without loss of generality, we assume that $VSS^*$ includes variable nodes (code-bits in this case) from both $G_U$ and $G_L$. We denote these two subsets of $VSS^*$ by $VSS_U^*$ and $VSS_L^*$, respectively.
Also, $GSS^*$ includes some variable nodes from the second column, i.e. from $v(1,2),..., v(N,2)$. Let us denote the index set of these nodes by $J^\prime$. For example, for the GSS shown in Fig. \ref{fig:stopset1}, $J^\prime$ is $\{2, 4, 6\}$.
We also denote the subsets of $J^\prime$ in the upper and lower halves of the graph by $J_U^\prime$ and $J_L^\prime$, respectively.
Furthermore, We simply use $T^U$ and $T^L$ instead of $T_{k+1}^U$ and $T_{k+1}^L$, since it is clear that we are dealing with the case $n=k+1$.
Accordingly,  we use $f_U(j^\prime)$  ($f_L(j^\prime)$)  to show the size of the leaf set for the stopping tree of $j^\prime \in J_U^\prime$ ($j^\prime \in J_L^\prime$) in $T^U$ ($T^L$).

For this setting, we need to show that for bit $i$ to be erased, at least $f(i)$ code-bits must be erased, or equivalently, $|VSS^*|\geq f(i)$. We consider two cases: 1. $i \in G_L$, and 2. $i\in G_U$.

\begin{enumerate}
  \item \emph{$i\in G_L$}:
This case is depicted in Fig \ref{fig:stopset3}. First, note that $i-2^k$ can not be in the $VSS^*$, because $f(i-2^k)=1/2f(i)$ and then $i$ would not be a MIB.
Now, for $i$ to be erased, $i^\prime$ and $l^\prime=i^\prime-2^k$ must be erased.
Fact \ref{fact:stopinduce} asserts that $J$ induces two stopping sets in $T^U$ and $T^L$ for $J_U^\prime$ and $J_L^\prime$, respectively.
We claim that $i^\prime$ and $l^\prime$ are MIB for $J_L^\prime$ and $J_U^\prime$, respectively. If $i^\prime\neq MIB(J_L^\prime)$, then there exists a node $j^\prime$ such that $f_L(j^\prime)<f_L(i^\prime)$. Then, there exists $j\in \mathcal{A}$ such that $f(j)<f(i)$ which is in contradiction with the fact that $i$ is a MIB.

If $l^\prime\neq MIB(J_U^\prime)$, then there exists $t^\prime$ such that $f_U(t^\prime)<f_U(l^\prime)$. This means that we have $t\in J$ and/or $t+2^k\in J$. However, we then have $f(t)<f(i)$ and $f(t+2^k)<f(i)$, which is again a contradiction with $i$ being a MIB. Now, since $i^\prime=MIB(J_L^\prime)$ and $l^\prime=MIB(J_U^\prime)$, then the induction hypothesis implies that $|VSS_L^*|\geq f_L(i^\prime)$ and $|VSS_U^*|\geq f_U(l^\prime)$. Therefore, $|VSS^*|=|VSS_L^*|+|VSS_U^*|\geq f_L(i^\prime)+f_U(l^\prime)=f(i)$.

  \item \emph{$i\in G_U$}: This case is depicted in Fig. \ref{fig:stopset4}.
If $J\cap G_L=\phi$, then we can prove that $i^\prime=MIB(J_U^\prime)$ along the same lines as the proof of case 1 above. Then the induction hypothesis implies that $VSS^*\geq f_U(i^\prime)=f(i)$, and the proof would be complete for this case.

Now suppose that $J\cap G_L\neq\phi$. Consider any $j\in J\cap G_L$. We show that $f(j)>f(i+2^k)$. Let us denote $i+2^k$ by $t$.
First note that $f(j)>f(i)$; otherwise if $f(j)=f(i)$, then according to our definition of MIB, we would pick $j$ as the MIB since $j\in G_L$ and $i\in G_U$.
Also note that $f(.)$ only takes value as powers of 2. Hence, we have $f(j)\geq 2f(i)$. Therefore, $f_L(j^\prime)=1/2f(j)\geq f(i)=f_L(t^\prime)$. As a result, $|VSS^*|\geq |VSS_L^*|\geq f_L(t^\prime)=f(i)$. $\blacksquare$
\end{enumerate}

\vspace{.1 in}

\emph{Proof of Fact \ref{fact:sizedist}:}
The fact becomes clear by looking at the recursive structure of the graph: $T_{n+1}$  is formed of two copies of $T_n$, one at the top and one at the bottom, that are connected together.
$\blacksquare$

\vspace{.1 in}

\emph{Proof of Theorem \ref{th:vanish}:
}In the matrix $F^{\otimes n}$, there are ${n\choose i}$ rows with weight $2^i$ \cite{hussami09}. This means that in the factor graph of a polar code, there are ${n\choose i}$ stopping trees with a leaf set of size $2^i$. Thus the corresponding tree of these input bits is at least of size $2^i$. As a result, the number of input bits with less than $2^{\epsilon n}=N^\epsilon$ variable nodes in their tree is less than $\sum_{i=0}^{\epsilon n}{n\choose i}$, which is itself upper-bounded by $2^{H(\epsilon) n}=N^{H(\epsilon)}$ for $0<\epsilon<\frac{1}{2}$. $\blacksquare$

\vspace{.1 in}

\emph{Proof of Theorem \ref{th:stopasymp}:}
The block error probability for SC decoding over every B-DMC is proved to be $O(2^{-\sqrt{N}})$  \cite{ArikanTelatar09}.
Noting that the error correction performance of BP is at least as good as SC over the BEC \cite{hussami09}, we conclude that block error probability for BP over the BEC decays as $O(2^{-\sqrt{N}})$ as well. Let us denote by $P_B(E)$ and $\textrm{Pr}\{E_{MVSS}\}$, the block erasure probability and the probability of MVSS being erased. We then have
\begin{align}
\textrm{Pr}\{E_{MVSS}\}=\epsilon^{|MVSS|}=(1/\epsilon)^{-|MVSS|} \leq P_B(E) = O(2^{-\sqrt{N}}) \Rightarrow |MVSS|=\Omega(\sqrt{N}),
\end{align}
where $\epsilon$ is the channel erasure probability. $\blacksquare$

\vspace{.1 in}

\emph{Proof of Theorem \ref{th:stopdmin}:
}First note that according to Fact \ref{fact:weight}, $f(i)=wt(\textbf{r}_i)$ for any $i\in \mathcal{I}$.
On the other hand, according to \cite{hussami09, Hassaniratedep11}, $d_{min}=\min_{i\in \mathcal{A}} wt(\textbf{r}_i)$ for a polar code. Now using Corollary \ref{cor:minVSS}, $d_{min}=\min_{i\in \mathcal{A}} wt(\textbf{r}_i) =\min_{i\in \mathcal{A}} f(i)=|MVSS|$. $\blacksquare$

\newpage


\end{document}